\documentclass[conference]{IEEEtran}
\IEEEoverridecommandlockouts
\usepackage{cite}
\usepackage[font=footnotesize]{caption}
\usepackage{amsmath,amssymb,amsfonts}

\usepackage{algorithmic,bm}
\usepackage[ruled,vlined]{algorithm2e}
\usepackage{mathtools}

\usepackage{graphicx}
\usepackage{textcomp}
\usepackage{xcolor,url}
\expandafter\def\expandafter\UrlBreaks\expandafter{\UrlBreaks%  save the current one
	\do\a\do\b\do\c\do\d\do\e\do\f\do\g\do\h\do\i\do\j%
	\do\k\do\l\do\m\do\n\do\o\do\p\do\q\do\r\do\s\do\t%
	\do\u\do\v\do\w\do\x\do\y\do\z\do\A\do\B\do\C\do\D%
	\do\E\do\F\do\G\do\H\do\I\do\J\do\K\do\L\do\M\do\N%
	\do\O\do\P\do\Q\do\R\do\S\do\T\do\U\do\V\do\W\do\X%
	\do\Y\do\Z}

\usepackage[font=footnotesize]{subcaption}
\begin{document}

\title{MagBB: Wireless Charging for Batteryless Sensors using Magnetic Blind Beamforming \\
}
\author{\IEEEauthorblockN{Albert Aninagyei Ofori and Hongzhi Guo}
\IEEEauthorblockA{{Engineering Department} \\
{Norfolk State University, Norfolk, VA, USA}\\
a.a.ofori@spartans.nsu.edu, hguo@nsu.edu}}

\maketitle

\begin{abstract}
Tiny batteryless sensors are desirable since they create negligible impacts on the operation of the system being monitored or the surrounding environment. Wireless energy transfer for batteryless sensors is challenging since they cannot cooperate with the charger due to the lack of energy. In this paper, a Magnetic Blind Beamforming (MagBB) algorithm is developed for wireless energy transfer for batteryless sensors in inhomogeneous media. Batteryless sensors with randomly orientated coils may experience significant orientation losses and they may not receive any energy from the charger. MagBB uses a set of optimized current vectors to generate rotating magnetic fields which can ensure that coils on batteryless sensors with arbitrary orientations can receive sufficient voltages for charging. It does not require any information regarding the batteryless sensor's coil orientation or location. The efficiency of MagBB is proven by extensive numerical simulations.      
\end{abstract}

\section{INTRODUCTION}

Wireless sensor plays an important role in precision agriculture. It can be used to collect information such as light intensity and leaf color. More recently, underground wireless sensors are also introduced to precision agriculture to monitor soil nutrition and moisture \cite{dong2013autonomous,vuran2018internet}. Today's wireless sensors are bulky and expensive, which is not suitable for large-scale wireless sensor networks. Hence, wireless sensors are used sparsely on farms. 

\begin{figure}[t]
	\centering
	\includegraphics[width=0.4\textwidth]{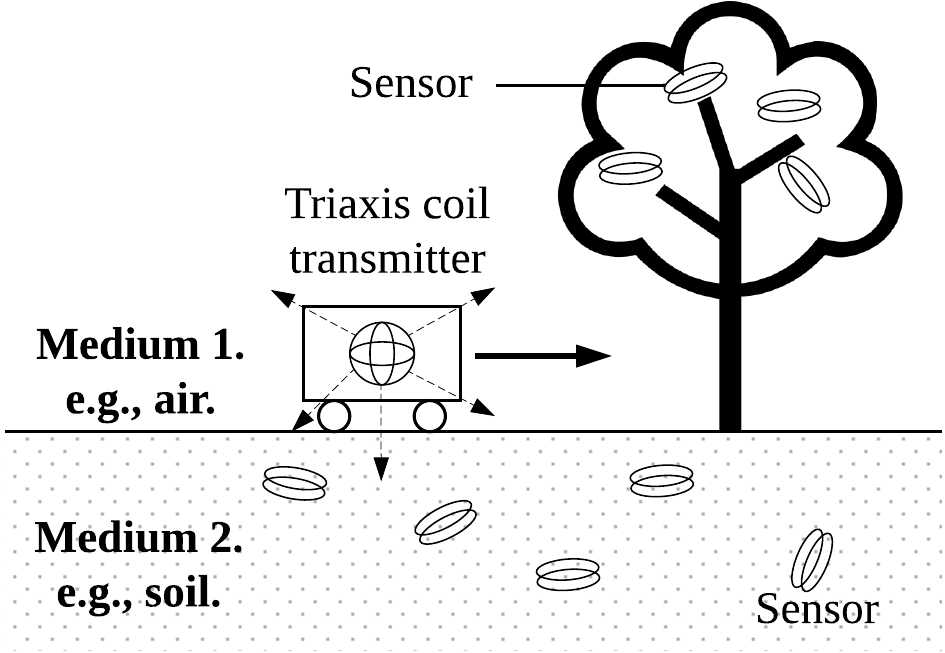}
	\vspace{-5pt}
	\caption{Illustration of a tri-axis coil in a moving robot being used to charge multiple batteryless sensors with tiny unidirectional coils. The batteryless sensors are used for precision agriculture. }
	\vspace{-10pt}
	\label{fig:1}
\end{figure}

However, as sensor technology improves, sensors can be made smaller and less intrusive thereby making them more suitable for in-situ sensing \cite{maselli2019battery,yunfei2018miniature}. Batteries are inconvenient in applications where sensors are not easily accessible as having to replace batteries, for instance, in an underground sensor would require precise knowledge of the sensor location and having to dig it up each time. Thus, tiny batteryless sensors are desirable. For example, batteryless sensors can be deployed aboveground and underground to monitor various parameters regarding plant health, as shown in Fig.~\ref{fig:1}. Since these sensors are small and low-cost, they do not require precise deployment, which is more convenient compared with traditional sensors. 

Batteryless sensors rely on wireless energy transfer to support their normal operation. As shown in Fig.~\ref{fig:1}, a mobile wireless charger is used to wirelessly charge batteryless sensors. Since sensors are randomly located surrounding a plant, their orientations and locations are unknown. Also, the plant growth is a dynamic process, previous information of sensor locations and orientations may not be useful. Existing applications of wireless energy transfer calculate a channel model based on which the power transfer implementation takes into account how the transmitted signal is altered by noise and the nature of the channel \cite{jadidian2014MagneticMH}. The channel estimation approach, however, is only ideal for active sensors which can provide channel information feedback to the charger. In such a system, blind beamforming becomes a good method of overcoming such unpredictable circumstances. Using an efficient blind beamforming implementation, signals can be added constructively at the receiver without the knowledge of channel status information.   
 
 In this paper, we use magnetic induction-based wireless energy transfer to charge randomly located and orientated batteryless sensors \cite{jadidian2014MagneticMH,yang2017magnetic}. 
We use magnetic induction because its operating frequency is in the HF band which demonstrates a long skin depth. In this case, if the communication range is much shorter than the wavelength, the impact of the soil-air boundary can be neglected \cite{guo2020performance} and, thus, we can consider the environment as a homogeneous medium. The mobile charger employs a tri-axis coil with three mutually perpendicular unidirectional coils. By optimally controlling the currents of the tri-axis coil, we can create magnetic fields in any direction. We design the Magnetic Blind Beamforming (MagBB) algorithm to rotate the magnetic field at a certain point to ensure that an arbitrarily orientated coil can receive sufficient power to charger the batteryless sensor. We compare our approach with other solutions to show its effectiveness and evaluate its performance numerically. 

The rest of this paper is organized as follows. In Section II, we discuss the research challenges and related works. In Section III, we present the MagBB algorithm and provide analytical insights to its performance. After that, in Section IV, the numerical analyses are given. Finally, this paper is concluded in Section V.

\section{Research Challenges and Related Work}

The wireless energy transfer for batteryless sensors in inhomogeneous media is challenging due to the following three reasons:
\begin{itemize}
	\item {\bf Threshold voltage.} Typical energy harvesting circuits use a full-wave bridge rectifier \cite{yunfei2018miniature} which consists of diodes to convert the induced AC voltages to DC voltages. Since diodes have a threshold voltage ${v}_{th}$, lower than which the circuit cannot harvest any energy. The typical $v_{th}$ is around 200~mV which is hard to obtain when the distance between a charger and a sensor is large. Moreover, considering the tiny size of the batteryless sensor, the antenna is not efficient in receiving energy due to its low profile. 
	\item {\bf Unknown coil/receiver information.} Since a batteryless sensor does not store any energy, it cannot perform channel estimation for wireless energy transfer. Without the channel status information, efficient beamforming cannot be implemented. Thus, the wireless charger has no information about the receiver. 
	\item {\bf Random coil/receiver location and orientation.} In an uncontrolled and relatively dynamic environment, the location of a sensor might change, or its antenna orientation in space might be altered by various factors. Even though sensors may be placed at specific, documented locations in the soil, the potentially dynamic nature of the internal soil environment can alter the exact location of a sensor over time. The growth of plants and the consequent penetration of their roots into the soil, for instance, has the potential to displace underground sensors to an unpredictable degree. Moreover, other uncontrollable phenomena such as erosion and seasonal changes in moisture can result in remarkable variations in communication channel conditions and the expected attenuation.
	\item {\bf Inhomogeneous environment.} The wireless signal propagation is hard to predict due to the inhomogeneous environment, which may require complex beamforming algorithms. However, it is not trivial to map the environment and predict signal propagation. 
\end{itemize}

The above challenges also arise individually in many other related works. In \cite{yunfei2018miniature}, the threshold voltage issue was addressed using blind beamforming analogous to the multipath fading in terrestrial environments. The UHF signals are used to power batteryless RFID sensors. However, the high carrier frequency may not efficiently penetrate through harsh environments that are considered in this paper. In \cite{yedavalli2017far,chen2016multiantenna}, adaptive blind beamforming is used for a multiantenna RFID reader. The RFID tag provides feedback signals to the reader to adaptively change the beamforming vectors. In this paper, the batteryless sensors do not have the capability to provide any feedback due to the lack of sufficient energy. Magnetic induction with long-wavelength signals can efficiently penetrate through dense media, which has been used for wireless communication in extreme environments \cite{kisseleff2014throughput,morag2019channel,guo2017multiple}. Also, the magnetic induction for wireless energy transfer has been widely accepted and the recently developed magnetic beamforming can significantly improve its efficiency \cite{jadidian2014MagneticMH, yang2017magnetic,zhao2020nfc+}. MagBB uses magnetic fields without knowing any information regarding the receiving coil, which is fundamentally different from existing solutions. 

\section{Magnetic Blind Beamforming using Tri-axis Coils}
In this section, we introduce the tri-axis coil model first. Then, we design a magnetic beamforming algorithm using current vectors. After that, we provide analytical analyses of the proposed algorithm and compare it with low-complexity solutions. Since we use long-wavelength magnetic induction signals, in the following we neglect the inhomogeneity of the environment and consider the sensors are located in homogeneous air medium. Also, due to the limited space of a batteryless sensor, it can only use a tiny unidirectional coil as the antenna. 
\subsection{Tri-axis Coil Array}
\begin{figure}
    \centering
    \includegraphics[scale=0.48]{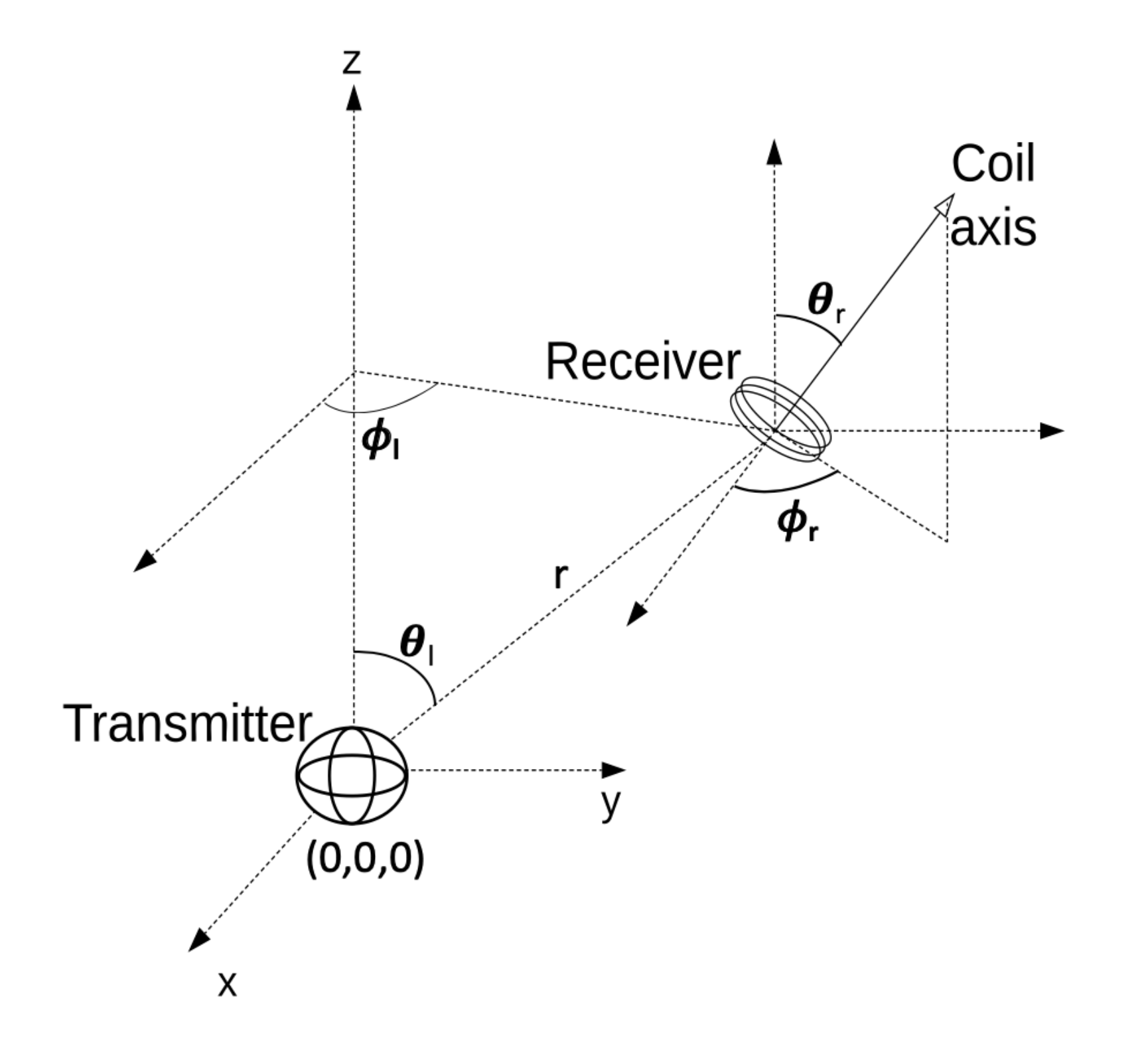}
        \vspace{-10pt}
    \caption{Illustration of the angular parameters used to describe the location and orientation of the receiver coil relative to the transmitter }
        \vspace{-20pt}
    \label{fig:2}
\end{figure}

The proposed approach to this problem is to design a tri-axis coil array whose magnetic field can be aligned in any direction in a three-dimensional (3D) space based on the value of the current that is fed into the individual coils. The current can be varied to produce a constantly rotating magnetic field that would be varied in fixed steps with a period that is long enough to provide enough energy to any sensor in a given direction to overcome its threshold voltage. The tri-axis coil, as shown in Fig.~\ref{fig:2}, has three coils with each coil axis aligned along each of the Cartesian plane axes (i.e., $x$, $y$, and $z$). 

We define pseudo-axes for a unidirectional coil using a unit vector ${\bm n}=[x_n, y_n, z_{n}]^T$. Since the coil size is much smaller than the wavelength, we consider a coil as a magnetic dipole and the dipole moment has the same direction as ${\bm n}$. To facilitate our analysis, we consider the dipole moment is created by three magnetic dipoles which only generate $x_n$, $y_n$, and $z_n$, respectively. The tri-axis coil can be considered as an array of three magnetic dipoles that creates arbitrary $x_n$, $y_n$, and $z_n$ by modulating the currents. Let the currents supplied to the coils with axes aligned along the $x_n$, $y_n$ and $z_n$ axis be $i_1$, $i_2$ and $i_3$. We assume the coils are identical and have the same radius $a_1=a_2=a_3=a_t$ and number of turns $N_1=N_2=N_3=N_t$. This simplifies the resultant formula for the magnetic field such that the total magnetic field observed at a point with spherical coordinates $(r, \theta_l, \phi_l)$ relative to the transmitter can be represented in spherical coordinates as ${\bm h}_r=[h_{rr}, h_{r\theta}, h_{r\phi}]^T$.
%\begin{table}
%	    \centering
%	
%	\caption{SIMULATION PARAMETERS}
%	\label{tab:1}
%
%\begin{tabular}{|c|c|c|c|}
%\hline
%     Original & Coil\textsubscript{3} & Coil\textsubscript{2}(Blue) & Coil\textsubscript{1}(Violet) \\
% \hline
% x  & x\textsubscript{3}  & y\textsubscript{2} & z\textsubscript{1}\\
%\hline
% y  & y\textsubscript{3}  & z\textsubscript{2} & x\textsubscript{1}\\
%\hline
% z  & z\textsubscript{3}  & x\textsubscript{2} & y\textsubscript{1}\\
% \hline
%\end{tabular}
%\end{table}

%Assuming circular coil n with its axis along the z-axis and is carrying a current $i\in\mathbb{C}$, the propagated magnetic field received at any point P at position $(r,\theta,\phi)$ from the origin in spherical coordinates, the magnetic field intensity $\bm{h}_n\in\mathbb{C}^3$ at this point due to the coil is\cite{guo2015channel}:
%
%\begin{equation}
%\bm{h}_n=\begin{bmatrix}
%    h_{rn}\\h_{\theta n}\\h_{\phi n}
%\end{bmatrix}=
%\begin{bmatrix}
%    j\frac{ka_n^2i_n\cos{\theta_n}}{2r^2}[1+\frac{1}{jkr}]e^{-jkr}\\
%    -\frac{(ka)^2i_n\sin{\theta_n}}{4r}[1+\frac{1}{jkr}-\frac{1}{(kr)^2}]e^{-jkr}\\0
%\end{bmatrix},
%\end{equation}

Since the aim is to achieve direction control by changing the currents, we can use one variable $\bm{i}\in\mathbb{C}^3$ to denote coil currents, i.e., ${\bm i}=[i_1, i_2, i_3]^T$. Then, the beamforming problem simplifies to the problem of obtaining optimal ${\bm i}$. To achieve this in a simplified manner, we introduce variables $C_r$, $C_{\theta}$, $C_{\phi}\in\mathbb{C}$ such that:
\begin{equation}\label{eq:4}
    \bm{C}=
    \begin{bmatrix}
        C_r\\C_{\theta}\\C_{\phi}
    \end{bmatrix}=
    \begin{bmatrix}
        \frac{jka^2_tN_t}{2r^2}\cdot[1+\frac{1}{jkr}]\cdot e^{-jkr}\\
        \frac{k^2a^2_tN_t}{4r} \cdot[1+\frac{1}{jkr}-\frac{1}{(kr)^2}]\cdot e^{-jkr}\\
        \frac{k^2a^2_tN_t}{4r}\cdot[1+\frac{1}{jkr}-\frac{1}{(kr)^2}]\cdot e^{-jkr}
    \end{bmatrix}
\end{equation}
where $k$ is the propagation constant and $a_t$ is the radius of the coil. Then, the magnetic fields generated by a tri-axis coil derived in \cite{guo2015channel} can be reformulated as 
\begin{align}
\label{eq:5}
\bm{h_r}=\bm{C}\cdot{\bm \Gamma}\cdot{\bm i}
\end{align} 
where
\begin{align}
    {\bm C}=\begin{bmatrix}
        C_r & 0 & 0\\
        0 & C_{\theta} & 0\\
        0 & 0 & C_{\phi}
    \end{bmatrix}
  \end{align}
  and 
  \begin{align}
  {\bm \Gamma}=
  \begin{bmatrix}
  \sin{\theta_l}\cos{\phi_l} & \sin{\theta_l}\sin{\phi_l} & \cos{\theta_l}\\
  \cos{\theta_l}\cos{\phi_l} & \cos{\theta_l}\sin{\phi_l} & -\sin{\theta_l}\\
  -\sin{\phi_l} & \cos{\phi_l} & 0
  \end{bmatrix},
  \end{align}  
where $\theta_l$ and $\phi_l$ are the angle parameters that are used to describe the location of the observation point, as shown in Fig.~\ref{fig:2}. Here, ${\bm C}$ is a function of distance, coil parameters, and environmental parameters and ${\bm \Gamma}$ is a function of the observation point location in the 3D space. For example, if receivers located on a sphere with the transmitter with a tri-axis coil in the center, they have the same ${\bm C}$ but different ${\bm \Gamma}$. 

Magnetic field ${\bm h}_r$ is obtained in a Spherical Coordinates System (SCS), while the coil orientation is described in a Cartesian Coordinates System (CCS). By using the following equation, we can obtain ${\bm h}_r$ in a CCS,
 \begin{equation}
\bm{h} = \bm{T} \cdot \bm{h_r},
\end{equation}
where 
\begin{equation}
    \bm{T} = \begin{bmatrix}
        \sin{\theta_l}\cos{\phi_l} & \cos{\theta_l}\cos{\phi_l} & -\sin{\phi_l} \\
        \sin{\theta_l}\sin{\phi_l} & \cos{\theta_l}\sin{\phi_l} & \cos{\phi_l} \\
        \cos{\theta_l} & -\sin{\theta_l} & 0
    \end{bmatrix}.
\end{equation}

 The aim is to produce an optimal magnetic field in the intended direction of observation and aligning the magnetic field vector in the direction of a unit vector along the axis of a coil at the observation point would be greatly desired. We define this unit vector as $\bm{u}\in\mathbb{R}^3$ such that:
 \begin{equation}
     \bm{u}=[\sin{\theta_r}\cos{\phi_r}, \sin{\theta_r}\sin{\phi_r},  \cos{\theta_r}]^T,
 \end{equation}
where $\theta_r$ and $\phi_r$ describe the orientation of the receiver coil axis, as shown in Fig.~\ref{fig:2}. 
To find three optimal current values can be complex in nature. This problem, can be modeled as an optimization problem that aims to reduce the distance between the resultant magnetic field and the unit vector in the direction of orientation of a coil at the observation point. We define a power threshold $P_{max}$ which should not be exceeded in supplying power to the antenna coils. We define a diagonal matrix $\bm{R}\in\mathbb{R}^{3\times3}$ representing the resistances of the tri-axis coil where $r_x$ is the resistance of transmitter coil $x$ and $r_1=r_2=r_3=r_t$ such that:
\begin{equation}
    \bm{R} = 
    \begin{bmatrix}
        r_t & 0 & 0\\
        0 & r_t & 0\\
        0 & 0 & r_t
    \end{bmatrix},
\end{equation}
where $r_t$ is the transmitter coil resistance.
The resultant magnetic field $\bm{h}$ should also induce a voltage large enough to overcome the threshold voltage in the receiver power harvesting circuit. The induced voltage $v$ can be expressed as:
\begin{equation}
    v=-\omega\mu_r\mu_0 N_r\pi a^2_r\bm{h\cdot u},
\end{equation}
where  $\mu_r$ is the relative permeability of the medium and $\mu_0$ is the permeability of free space, $N_r$ is the number of turns of the receiver coil and $a_r$ is the radius of the receiver coil.

\subsection{Magnetic Blind Beamforming}
To efficiently charge batteryless sensors, we need to maximize the induced voltage given limited transmission power. If we have the knowledge of the receiver's coil orientation ${\bm u}$, this problem can be efficiently solved. However, in this paper, we assume we have no knowledge of ${\bm u}$, thereby it is impossible to obtain the optimal ${\bm i}$. 

\begin{figure}
	\centering
	\begin{subfigure}[b]{.22\textwidth}
		\centering
		\includegraphics[width=\textwidth]{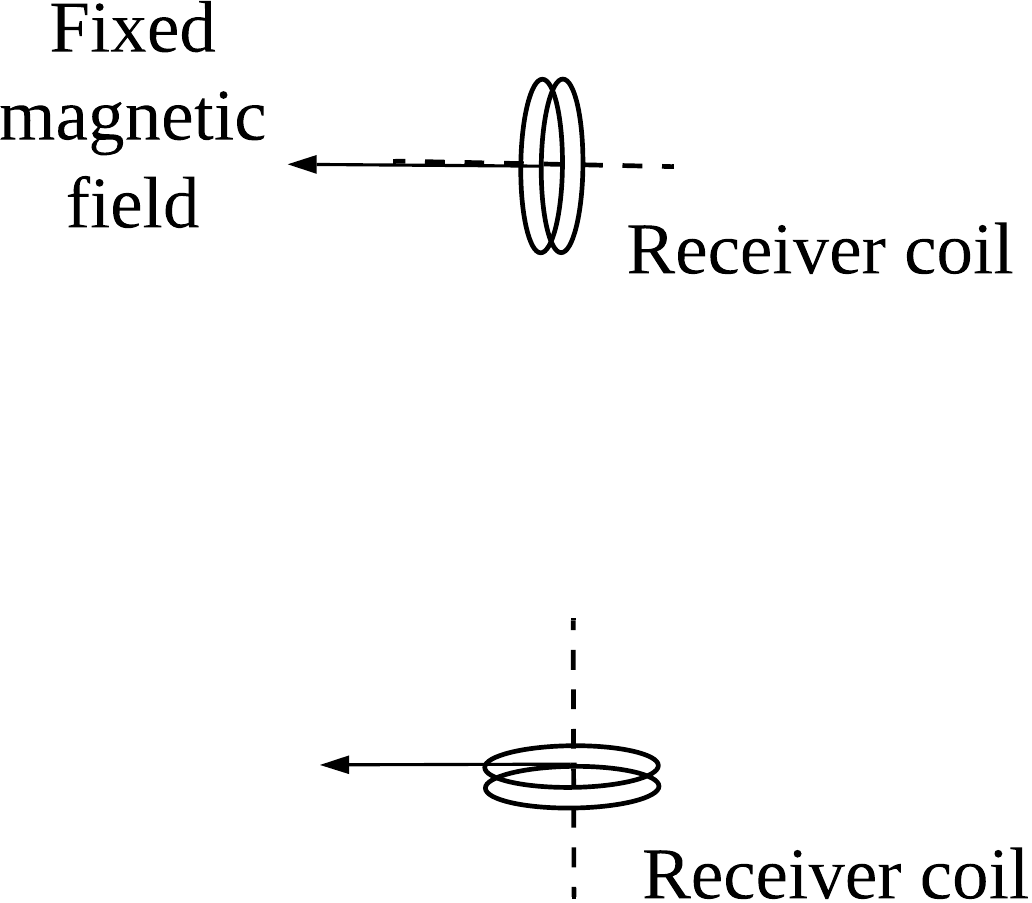}
		\caption{Constant transmitter current supply to charge randomly oriented coils}
		\label{fig:3a}
	\end{subfigure}
	\hfill
	\begin{subfigure}[b]{.22\textwidth}
		\centering
		\includegraphics[width=\textwidth]{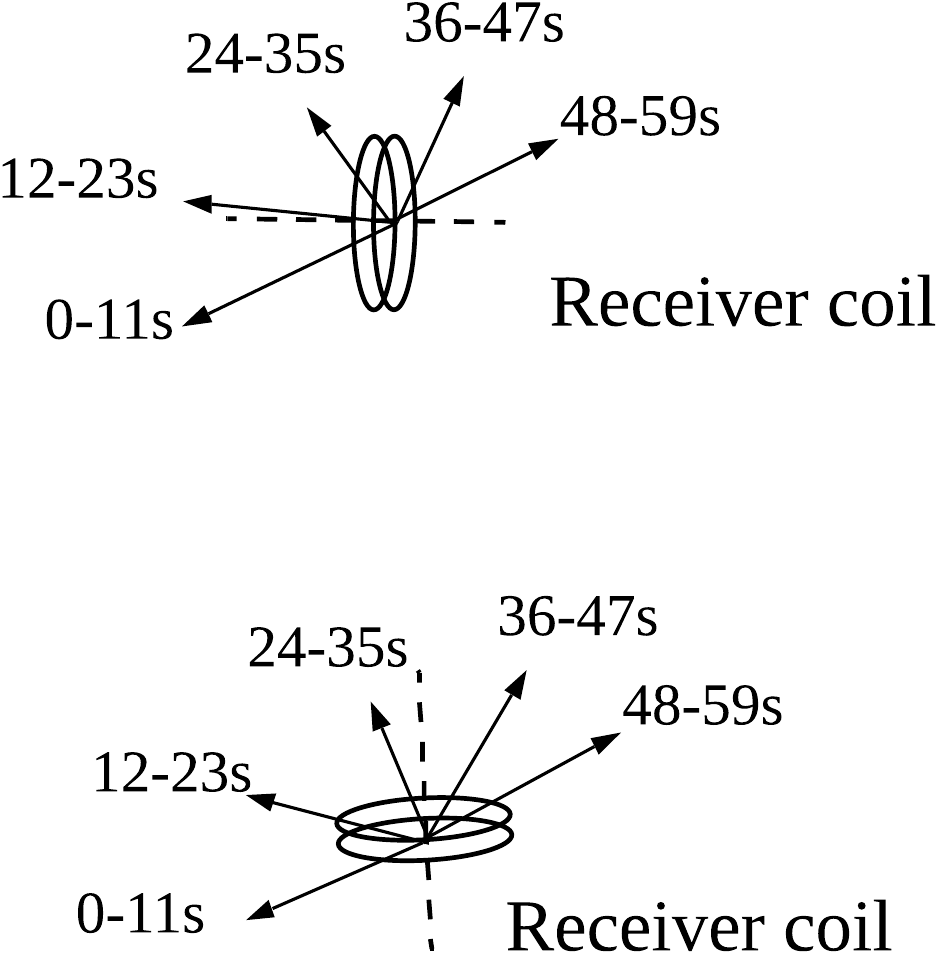}
		\caption{Transmitter current set rotating magnetic field through space.}
		\label{fig:3b}
	\end{subfigure}
	\caption{Illustration comparing performance of constant and rotating magnetic fields for different receiver coil orientations. Fig.~\ref{fig:3a} shows two instances  where maximum voltage is induced when coil axis is aligned with magnetic field (top) and zero voltage induced when coil axis is perpendicular to magnetic field (bottom).  Fig.~\ref{fig:3b} shows how voltage is induced in the receiver coil by the rotating magnetic field at some points during the charging cycle irrespective of coil orientation.} 
	\label{fig:3}
	\vspace{-10pt}
\end{figure}

To overcome the threshold voltage and efficiently charge coils with arbitrary orientations, we use MagBB and an example is shown in Fig.~\ref{fig:3}. First, when we use constant currents for tri-axis coils, the generated magnetic field always points to one direction. Different from electromagnetic wave-based wireless communication/charging in the terrestrial environments, the magnetic induction-based wireless communication/charging operates in the near field with a short distance, where the multipath fading can be neglected. As shown in the upper of Fig.~\ref{fig:3a}, if the magnetic field direction is parallel with the receiver's coil axis, the received power can be maximized. However, if the magnetic field direction is perpendicular to the receiver's coil axis, no voltage can be induced. Thus, the performance of using constant currents for the tri-axis coil is highly random, which is not predictable nor reliable without the knowledge of ${\bm u}$. 

MagBB is based on current vector rotation. As shown in Fig.~\ref{fig:3b}, 5 current vectors are used to create 5 magnetic fields with different directions (their magnitudes may also be different). For example, if we use 60 seconds to charge a sensor, each current vector is allocated with 12 seconds. Since the 5 vectors cover 180 degrees, we can ensure that coils with arbitrary orientations can be covered and at least one current vector can induce voltage that is larger than the threshold voltage to charge a sensor. Next, we try to study the following three problems: 1) how to design the current vectors? 2) how many current vectors do we need? and 3) does the current vector depend on the receiver's location? 

\subsubsection{Current Vector Design}
Assume we have the knowledge of the receiver's location. Then, we create $n_{cv}$ current vectors to rotate the magnetic field direction for 180 degrees. The following problem is formulated to obtain the current vector ${\bm i}$, 
\begin{align}
     \label{eq:10}
     \text{(P1):}~&\underset{\bm{i}\in\mathbb{C}^3}{\operatorname{minimize}} \left\lVert \bm{u}-\frac{\bm{h}}{\parallel  {\bm h}\parallel}\right\rVert^2\\
    &s.t.  ~~\bm{i^TR i}\leq 2P_{max};\\
&|\bm{h}^T{\bm u}|>\frac{v_{th}}{\omega\mu_r\mu_0 N_r\pi a^2_r},
\end{align}
where ${\bm u}$ is the targeting magnetic field direction. The objective function aims to design the magnetic field to point to the desired direction. The first constraint is due to the transmission power and the second constraint ensures the induced voltage is larger than the rectifier's threshold voltage $v_{th}$. Consider the targeting magnetic field directions as $\{{\bm u}_l, l=1, \cdots, n_{cv}\}$. By substituting $\{{\bm u}_l, l=1, \cdots, n_{cv}\}$ into the above problem, we can obtain the associated current vectors.

However, directly solving (P1) is challenging. In order to homogenize this optimization problem, a slack variable $t\in\mathbb{R}$ is introduced such that: 
\begin{equation}
    t=\parallel\bm{h}\parallel \text{ and } t^2=\bm{i}^T\cdot \bm{C}^T\cdot \bm{C}\cdot \bm{i}.
\end{equation}

The objective function can then be rewritten as 
\begin{align}
    \text{(P2): }&\underset{\bm{i}\in\mathbb{C}^3}{\operatorname{minimize}}  \parallel t\bm{u}-\bm{h}\parallel^2\\
    &s.t.~\bm{i}^T\bm{C}^T\bm{C}\bm{i}=t^2\\
    &~~~~\bm{i^TR i}\leq 2P_{max}\\
    &~~~~{\bm i}^T\bm{C}^T{\bm u}{\bm u}^T\bm{C}\bm{i}>\frac{V_{th}^2}{\omega^2(\mu_r\mu_0 N_r\pi a^2_r)^2} 
\end{align}
which can subsequently be solved using the method of  Semidefinite Relaxation \cite{luo2010semidefinite}.

Our ability to use Semidefinite Relaxation is incumbent on the ability to express our problem in the form ${\bm x}^T{\bm A}{\bm x}$ where $\bm{x}\in\mathbb{C}^N$ is the variable or system of variables to be optimized and $\bm{A}\in\mathbb{C}^{N\times N}$ is a function or group of functions with known values that constitute a positive semidefinite matrix. Since the value of $\bm{i}$ could be complex in nature, the optimization problem can be expressed as a real value equation whereby we decompose all potentially complex values into their real and imaginary forms such that:
\begin{align}
\bm{u}= &\begin{bmatrix}
    \Re\{\bm{u}\}\\
    \Im\{\bm{u}\}
\end{bmatrix},~~~
\bm{i}= \begin{bmatrix}
    \Re\{\bm{i}\}\\
    \Im\{\bm{i}\}
\end{bmatrix}\\
\bm{C}= &\begin{bmatrix}
    \Re\{\bm{C}\} & -\Im\{\bm{C}\}\\
    \Im\{\bm{C}\} & \Re\{\bm{C}\}
\end{bmatrix}\\ 
\bm{R}= &\begin{bmatrix}
    \Re\{\bm{R}\} & -\Im\{\bm{R}\}\\
    \Im\{\bm{R}\} & \Re\{\bm{R}\}
\end{bmatrix}    
\end{align}

This decomposition essentially doubles all dimensions such that all any matrix $\bm{X}\in\mathbb{C}^{N\times N}$ becomes $\bm{X}\in\mathbb{R}^{2N\times 2N}$ and all vectors $\bm{x}\in\mathbb{C}^N$ become $\bm{x}\in\mathbb{R}^{2N}$, i.e., $\bm{u}$ and $\bm{i}$ become  $6\times 1$ matrices and $\bm{C}$ becomes a $6\times 6$ matrix. The homogenized objective function $\parallel t\bm{u}-\bm{h}\parallel^2$ can then be accurately written as:

\begin{equation}
 \begin{bmatrix}
\bm{i}^T & t
\end{bmatrix}
{\bm A}
\begin{bmatrix}
\bm{i}\\ t
\end{bmatrix}=
    \begin{bmatrix}
    \bm{i}^T & t
    \end{bmatrix}
    \begin{bmatrix}
    \bm{C^TC} & \bm{-C^Tu}\\
    \bm{-u^TC} & \parallel \bm{u}\parallel^2
    \end{bmatrix}
    \begin{bmatrix}
        \bm{i}\\ t
    \end{bmatrix}
\end{equation}

In this form, we can adequately apply semidefinite programming in solving this problem. We can express a single variable $\bm{\beta}\in\mathbb{R}^{7\times 7}$ as the product of our variable matrix and its transpose:
\begin{equation}
\begin{bmatrix}
    \bm{i}\\
    t
\end{bmatrix}
\begin{bmatrix}
    \bm{i}^T & t
\end{bmatrix}=\bm{\beta}.
\end{equation}
The algorithm to be employed based on the objective function and constraints is given in Algorithm~\ref{alg:algorithm}.
\begin{algorithm}[t]
\SetAlgoLined
\caption{Code for optimization problem (P2)}
\label{alg:algorithm}
\begin{algorithmic}[1]
\STATE Variable $\bm{\beta}$(7,7) symmetric
\STATE $\text{minimize}(trace(\bm{A}*\bm{\beta}))$
\STATE subject to
\STATE \quad $trace(\bm{R}*\bm{\beta}\text{(1:6,1:6))}\leq 2P_{max}$
\STATE \quad $trace(\bm{C}^T\bm{C}*\bm{\beta}\text{(1:6,1:6))}= \bm{\beta}\text{(7,7)}$
\STATE \quad $trace((\bm{C}^T{\bm u}{\bm u}^T\bm{C})*\bm{\beta}
 \text{(1:6,1:6))}\geq \frac{V_{th}^2}{\omega^2(\mu_r\mu_0 N\pi a^2)^2}$
\end{algorithmic}
\end{algorithm}

The approximated solution is a rank-1 matrix $\bm{\beta}$ and to obtain the approximated ${\bm i}$ we use the dominant eigenvector:
\begin{equation}
    \bm{\beta} \approx \lambda_{max}{\bm q}{\bm q}^T,
\end{equation}
where $\lambda_{max}\in\mathbb{R}$ is the largest eigenvalue of matrix $\bm{\beta}\in\mathbb{R}$ and $\bm{q}\in\mathbb{R}^{N\times 1}$ is the eigenvector associated with $\lambda_{max}$. The eigenvector that is associated with the largest eigenvalue of $\bm{\beta}$ is considered as the solution ${\bm i}$.  From this approximation, the optimal decomposed current vector can be obtained as $\bm{i}^{\star}\in\mathbb{R}^{7\times 1}$ such that:
\begin{equation}
    \bm{i}^{\star}=\sqrt{\lambda_{max}}\cdot\bm{q},
\end{equation}
with 
\begin{align}
\begin{bmatrix}
\Re\{\bm{i}_1\}\\
\Re\{\bm{i}_2\}\\
\Re\{\bm{i}_3\}
\end{bmatrix} 
=\bm{i}^{\star}(1:3);~~\begin{bmatrix}
\Im\{\bm{i}_1\}\\
\Im\{\bm{i}_2\}\\
\Im\{\bm{i}_3\}
\end{bmatrix}
=\bm{i}^{\star}(4:6).
\end{align}

The performance of a given set of currents is then obtained by finding the optimum current for a coil in a given location and in all possible directions of orientation. The efficacy of this system lies in its ability to induce a voltage in a randomly oriented coil at a given location for a sufficiently long period such that the induced voltage $v>v_{th}$.

\subsubsection{Number of Current Vectors}

We cycle through the current set in a fast enough manner in order to keep the sensors charged. For a given set of current values, therefore, we charge the sensor long enough such that all the currents in the current set are applied for an equal time interval of $\Delta T$, i.e.,
\begin{equation}\tag{19}
    \Delta T\times n_{cv} =T_c,
\end{equation}
where $T_c$ is the overall charging time. Here, there is a trade-off between the charging efficiency and the charging time interval of $\Delta T$. On one hand, if $n_{cv}$ is large, with high probability, we can obtain the optimal (or nearly optimal) current vector to charge the receiver with near-zero orientation loss. On the other hand, a large $n_{cv}$ reduces $\Delta T$ since the $T_c$ is a constant. Even if the charging efficiency is high, a small $\Delta T$ may reduce the overall received energy. Since the receiver's orientation ${\bm u}$ is an arbitrary vector in 3D space, to fully capture all possibilities, we need three current vectors at least. Otherwise, if $n_{cv}<3$, there are some blind points. On the contrary, there is no limitation on the maximum number of current vectors since with more current vectors, we can cover more orientations. However, as the number increases, the complexity of the algorithms increases, and the correlation between two adjacent current vector increases, which may waste resources. 

The minimum number of current vectors, i.e., 3 mutually orthogonal receiver coil orientations, can be designed in the following way. We define three random vectors $\bm{v_1, v_2, v_3} \in\mathbb{R}^{3\times 1}$ based on which we derive orthogonal basis $\bm{u_1, u_2, u_3} \in\mathbb{R}^{3\times 1}$ using Gram–Schmidt process such that:
\begin{align}
    \bm{u_1}= & \bm{v_1}\\
    \bm{u_2}= & \bm{v_2}-\frac{\bm{v_2}\cdot \bm{u_1}}{\lVert \bm{u_1}\rVert^2}\bm{u_1}\\
    \bm{u_3}= & \bm{v_3}-\frac{\bm{v_3}\cdot \bm{u_1}}{\lVert \bm{u_1}\rVert^2}\bm{u_1} - \frac{\bm{v_3}\cdot \bm{u_2}}{\lVert \bm{u_2}\rVert^2}\bm{u_2}
\end{align}
By normalizing $\bm{u_1}$, ${\bm u_2}$, and ${\bm u_3}$, we obtain the orthonormal basis $\bm{e_1, e_2, e_3} \in\mathbb{R}^{3\times 1}$ which represent three mutually orthogonal coil directions such that:
\begin{equation}
    e_n=\frac{\bm{u_n}}{\lVert\bm{u_n}\rVert}
\end{equation}
Knowledge of these orientations can then be used in finding the optimum currents for a coil aligned along these directions by substituting into \eqref{eq:10}.

\subsubsection{Receiver Location Dependency}
Note that, we optimize the current vectors at one specific location. Next, we explore how the location of a receiver coil affects the magnetic field intensity, and how to maximize this field. From \eqref{eq:5}, we introduce the scalar and angular coefficients of the equation. It can be seen that $\bm{\Gamma}$ is an orthogonal matrix since $\bm{\Gamma}\bm{\Gamma}^T=\bm{\Gamma}^T\bm{\Gamma}=\bm{I}$. The angular coefficient, therefore, does not affect the magnitude of the generated magnetic field vector but would only either reflect it in some plane or rotate it. The factor affecting the efficacy of a given current vector set in any location would, therefore, depend principally on the scalar coefficient $\bm{C}$. From \eqref{eq:4}, it can  be seen that $C_{\theta}=C_{\phi}$. Also, it can be seen, quite intuitively, that the scalar coefficient magnitudes are inversely proportional to $r$. In the near field of the coil antenna ($r\ll \text{wavelength}/2\pi$),
\begin{equation}
\label{equ:20}
   C_r\approx 2C_{\theta}\approx 2C_{\phi}.
\end{equation}
The induced voltage is proportional to ${\bm h}_r$ which is given in \eqref{eq:5}. Since ${\bm \Gamma}$ is an orthogonal matrix, ${\bm h}_r$ is equivalent to the result of projecting ${\bm i}$ in a new orthogonal coordinates system using ${\bm \Gamma}$ and then scaled the three elements using $C_r$, $C_{\theta}$, and $C_{\phi}$. Given the distance, $C_r$, $C_{\theta}$, and $C_{\phi}$ do not change on the sphere. In the near field, due to \eqref{equ:20}, ${\bm h}_r$ are similar, which does not depend on the location on the sphere. 
Assuming the applied current is at its maximum accepted value such that $\lVert\bm{i}\rVert^2_{\text{max}} = \frac{2P_{\text{max}}}{r_t}$,
then 
${C_{\theta}}\lVert \bm{i}\rVert_{\text{max}}\leq \lVert\bm{h}\rVert\leq{C_{r}}\lVert \bm{i}\rVert_{\text{max}}$,
and this stands true for the near-field region. This relatively low variation in magnetic field intensity as the receiver location changes for a constant current is very convenient for reliable charging of a receiving sensor.  

However, as we venture into the far field ($kr>>1$)\cite{balanis2016antenna}, the scalar coefficient values start to diminish such that $C_r<<C_{\theta}\approx C_{\phi}$. In this case, the projected ${\bm i}$ is scaled unevenly. If the projected vector has a dominant element and it is scaled by $C_r$, the ${\bm h}_r$ can be extremely small. The magnitude of the magnetic field can, therefore, vary widely as the receiver location changes, potentially resulting in erratic charging efficiency as the produced voltage can drop below the desired value with a slight change in the relative location of the receiver to the transmitter. This makes a far field application of this implementation relatively suboptimal. 

\section{SIMULATION AND RESULTS}
In this section, various simulations are carried out to numerically analyze, compare, and contrast the performance of the blind beamforming scheme under various conditions. We first evaluate the induced voltage and then we generate randomly orientated receiving coils and compare the reliability of different approaches. The employed values for the simulations are shown in Table~\ref{tab:2}.

\begin{table}[t]
	\centering
	
	\caption{SIMULATION PARAMETERS}
	\label{tab:2}
	
	\begin{tabular}{|c|c||c|c|}
		\hline
		Quantity & Value & Quantity & Value \\
		\hline\hline
		$r$  & $0.6$ m / $1.2$ m  & $f$ & $13.56$ MHz\\
		\hline
		$\epsilon_o$  & $8.85\times 10^{-12}$ F/m  & $\epsilon_r$ & 1.0006\\
		\hline
		$\mu_o$ & $1.2566\times10^{-6}\text{mkgs}^{-2}\text{A}^{-2}$ & $\mu_r$ & 1\\
		\hline
		$a_t$ & $0.1$ m  & $a_r$ & $0.01$ m\\
		\hline
		$P_{\text{max}}$ & $50$ W & $v_{\text{th}}$ & $0.2$ V\\
		\hline
		$r_t$ & $1~\Omega$ & $r_r$ & $0.2~\Omega$\\ 
		\hline
	\end{tabular}
\end{table}

\subsection{Comparison of Induced Voltage}
\begin{table}[t]
	\centering
	
	\caption{LOCATION PARAMETERS}
	\label{tab: location params}
	
	\begin{tabular}{|c|c|c|}
		\hline
		\textbf{Simulation Locations} & $\bm{\theta_l}$ & $\bm{\phi_l}$ \\
		\hline\hline
		Optimized location  & $180^{\circ}$ & $0^{\circ}$ \\
		\hline
		Random location 1  & $0.021^{\circ}$  & $108.84^{\circ}$ \\
		\hline
		Random location 2 & $33.53^{\circ}$ & $124.4^{\circ}$ \\
		\hline
	\end{tabular}
\end{table}

\begin{figure}[t]
	\centering
	\includegraphics[width=8cm,height=6cm]{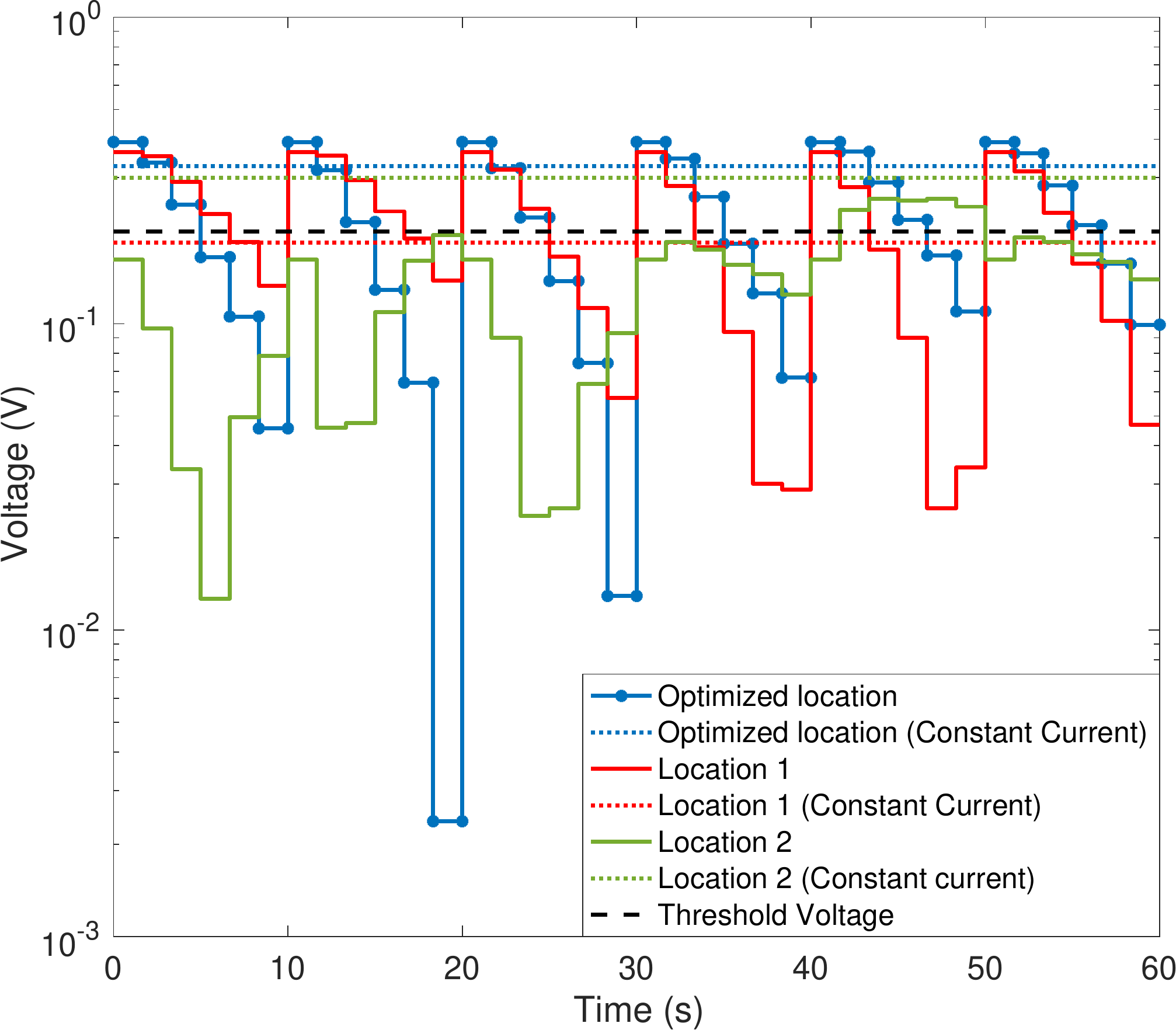}
	\caption{Performance comparison for 36 current sets.}
	\label{fig:4}
\end{figure}
\begin{figure}
	\centering
	\includegraphics[width=8cm,height=6cm]{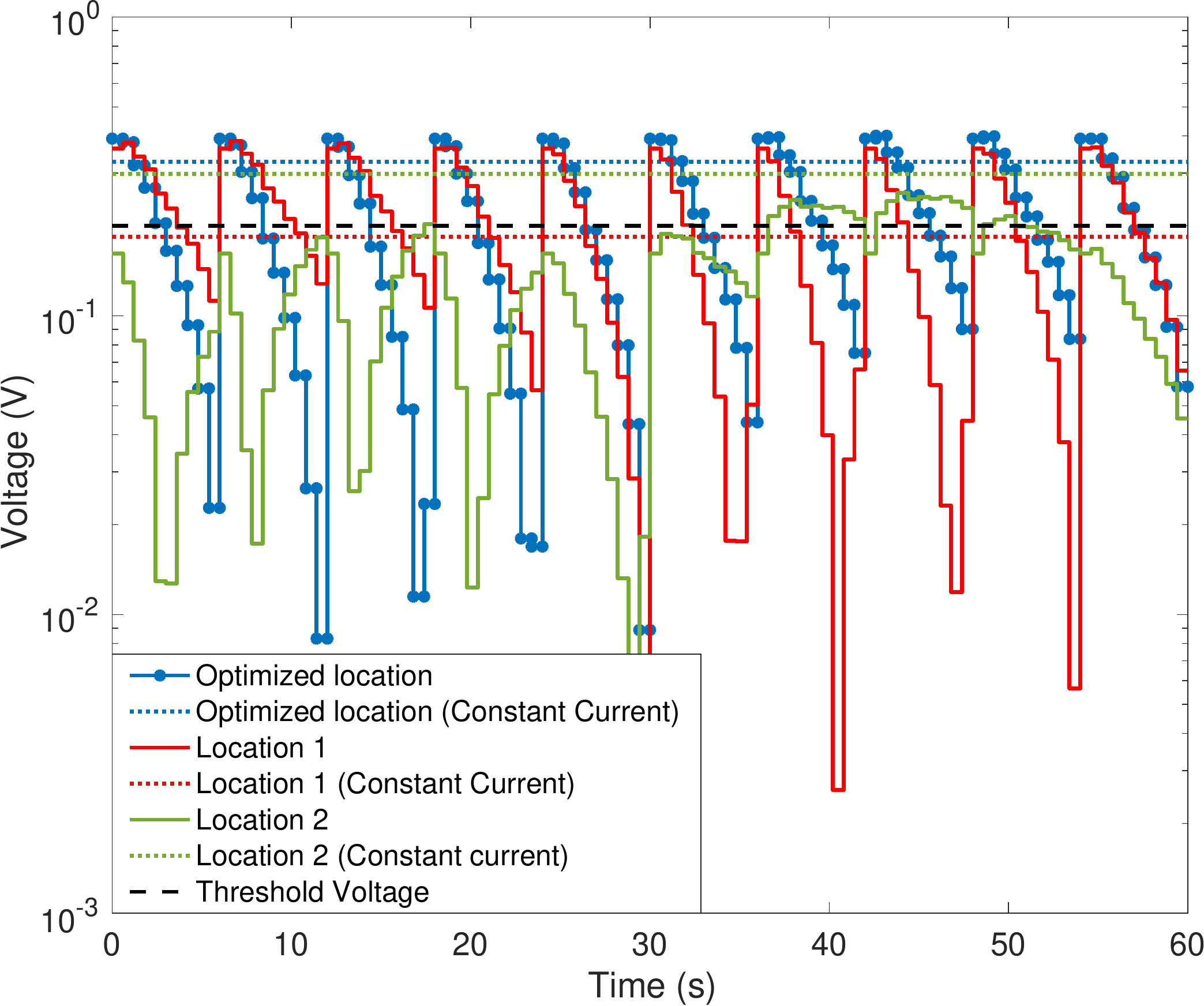}
	\caption{Performance comparison for 100 current sets.}
	\label{fig:5}
\end{figure}

We test the performance of this system using a current set optimized for randomly oriented coil at a predetermined location 1.2~m directly below the transmitter ($\theta_l=180^{\circ}$, $\phi_l=0^{\circ}$), which is called the ``Optimized location'' in Fig.~\ref{fig:4} to Fig.~\ref{fig:9}. Also, we compared with two other locations given in Table~\ref{tab: location params}, which are called ``Location 1'' and ``Location 2'', respectively, in Fig.~\ref{fig:4} to Fig.~\ref{fig:9}. We then rotated the various coil orientations throughout the spherical region in predetermined steps and calculate the corresponding current values based on the optimization method outlined in Algorithm~\ref{alg:algorithm}. We compare the induced voltage at a receiver placed at this location by the calculated current set to that for 2 other randomly oriented coils located at randomly determined locations as shown in Table~\ref{tab: location params}. Lastly, we employ a simplistic approach by finding the maximum balanced current without exceeding the power threshold. We calculate the induced voltage at the predetermined locations and compared these with the results of the optimal current sets. Since the circuitry can be configured to invert negative voltages, the working range of the sensor consists of all induced voltages $v$ such that
$ |v|> v_{th}$.
We compare the voltage induced within a one-minute charging cycle, and the total energy induced for various radial distances, locations, and orientations as outlined in the following sections.

We compare two optimized current sets; we calculate one set of currents for 36 combinations of $\theta_r$ and $\phi_r$ throughout a hemispherical range. We repeat this process for a second set of currents for 100 locations within the hemispherical range, therefore, giving a better resolution. To test the efficacy of these current sets, we find the voltage induced for randomly located and oriented coils at a distance of 1.2~m from the transmitter by charging the receiver coils for an equal amount of time by each current set within a 1-minute charging cycle. We also use the simplistic approach by supplying the maximum allowable current to the transmitter coils per the power requirements specified for this system. i.e.,
\begin{align}
 \bm{i}=\sqrt{\frac{2P_{\text{max}}}{3r_t}}[1,1,1]^T
\end{align}

The performance of the optimized and simplistic approaches in charging multiple randomly located and oriented receivers is compared. Fig.~\ref{fig:4} and Fig.~\ref{fig:5} show this comparison for the 36-current and 100-current sets respectively. The optimized set shows a clear advantage here as it is able to charge all 3 randomly located receivers above the threshold voltage at certain points within the charging cycle. The constant current, however, either works throughout the charging cycle (optimized location  \& location 2) or does not charge the receiver at all throughout the charging cycle (location 1). This, therefore, makes the simplistic approach suboptimal for simultaneous charging at multiple locations.  Comparing Fig.~\ref{fig:4} and Fig.~\ref{fig:5}, the 100-current set appears to charge the receiver more continuously given the higher resolution afforded by more current sets and more continuous rotation in the three-dimensional space. 

However, this advantage comes with more computational burden given the higher number of steps required. We explore the necessity and possibility of reducing this computational burden by producing a set of current vectors optimized for three mutually orthogonal coil orientations which are then tested in comparison with the constant current sets for the same locations as shown in Table~\ref{tab: location params}.
In order to observe the performance of the system with a minimum number of current values, we determine the orthonormal bases of 3 randomly generated unit currents. The optimum current
values are then found for each orientation using the Algorithm~\ref{alg:algorithm}. The performance is observed as shown in Fig.~\ref{fig:6} and Fig.~\ref{fig:9}. As we can see, the 3 orthogonal current vectors can also achieve high induced voltage. Also, it provides more directional diversity compared with the constant current. 
\subsection{Reliability Evaluation}
We then analyze the performance of the various current sets for fixed locations at 0.6~m and at 1.2~m. The general performance is appraised by finding the total energy delivered to 10,000 randomly positioned receivers assuming the receivers are only charged by voltages above the threshold voltage. The receiving coils are randomly orientated by letting their axes point to a random point on a unit sphere. Then, we calculate the received energy during one charging cycle. To make a fair comparison, we consider one cycle is 60~s, and a current vector uses $60/n_{cv}$~s. Then, we obtain the CDFs (Cumulative Distribution Function) for various policies and compare their performances. 

\begin{figure}
	\centering
	\includegraphics[width=8cm, height=6cm]{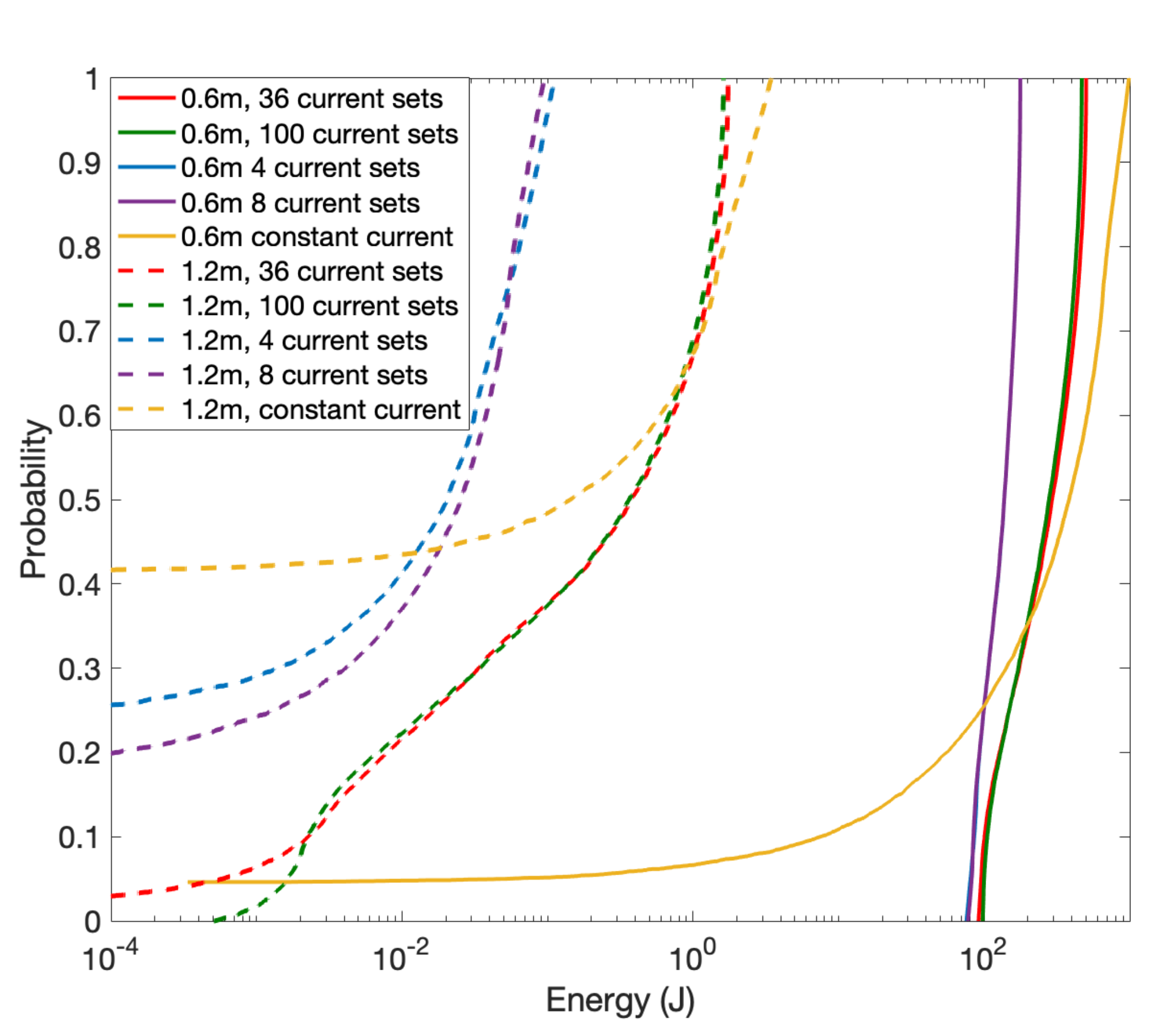}
	\caption{ CDF comparing different orientations of different current sets for fixed location $\theta_l=180^{\circ}$ and $\phi_l=0^{\circ}$ at a distance of 0.6~m and 1.2~m}
	\label{fig:8}
\end{figure}

\begin{figure}
	\centering
	\includegraphics[width=8cm, height=6cm]{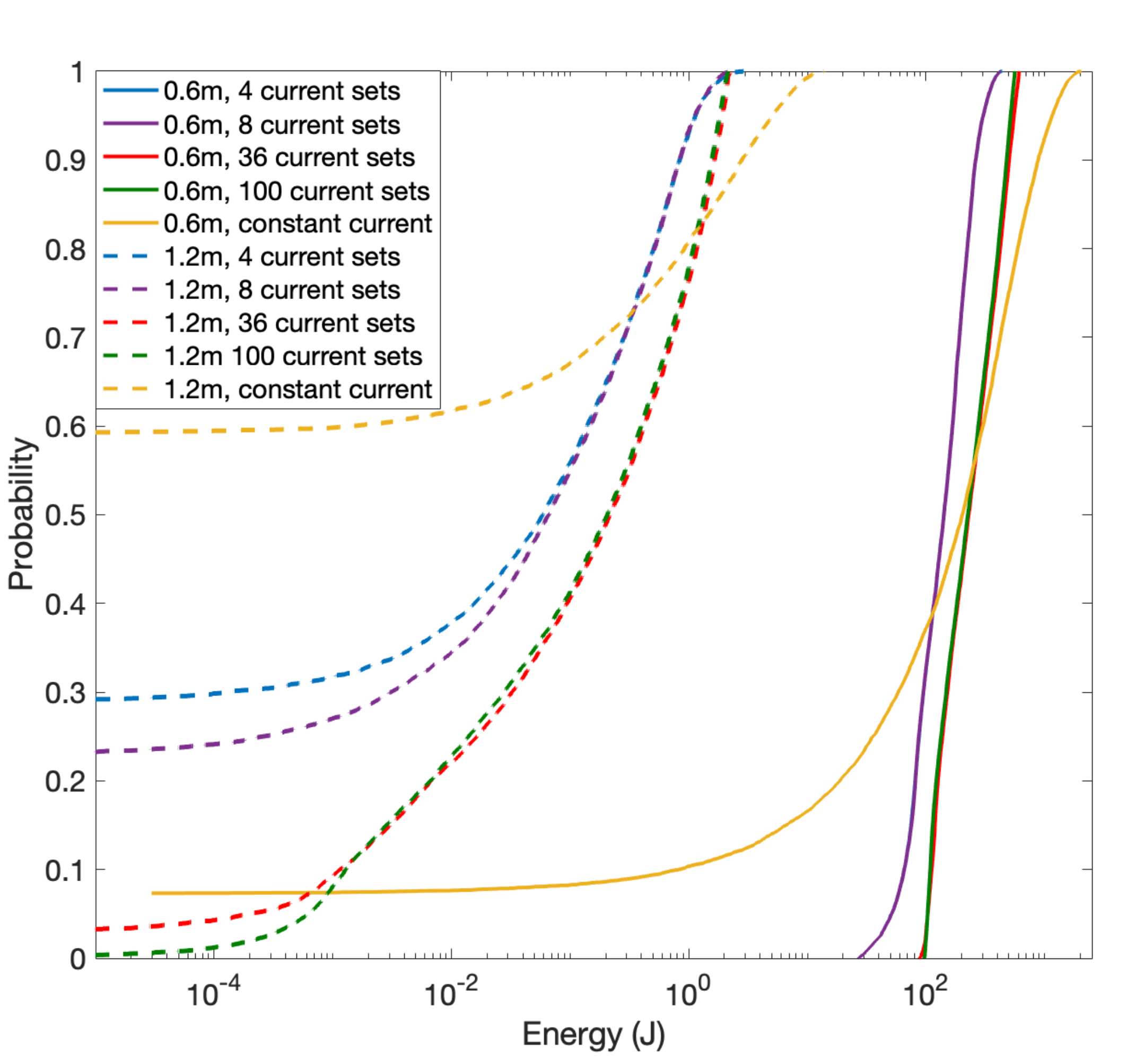}
	\caption{CDF for random locations and orientations in 3D space}
	\label{fig:7}
\end{figure}

\begin{figure}
	\centering
	\includegraphics[width=8cm, height=6cm]{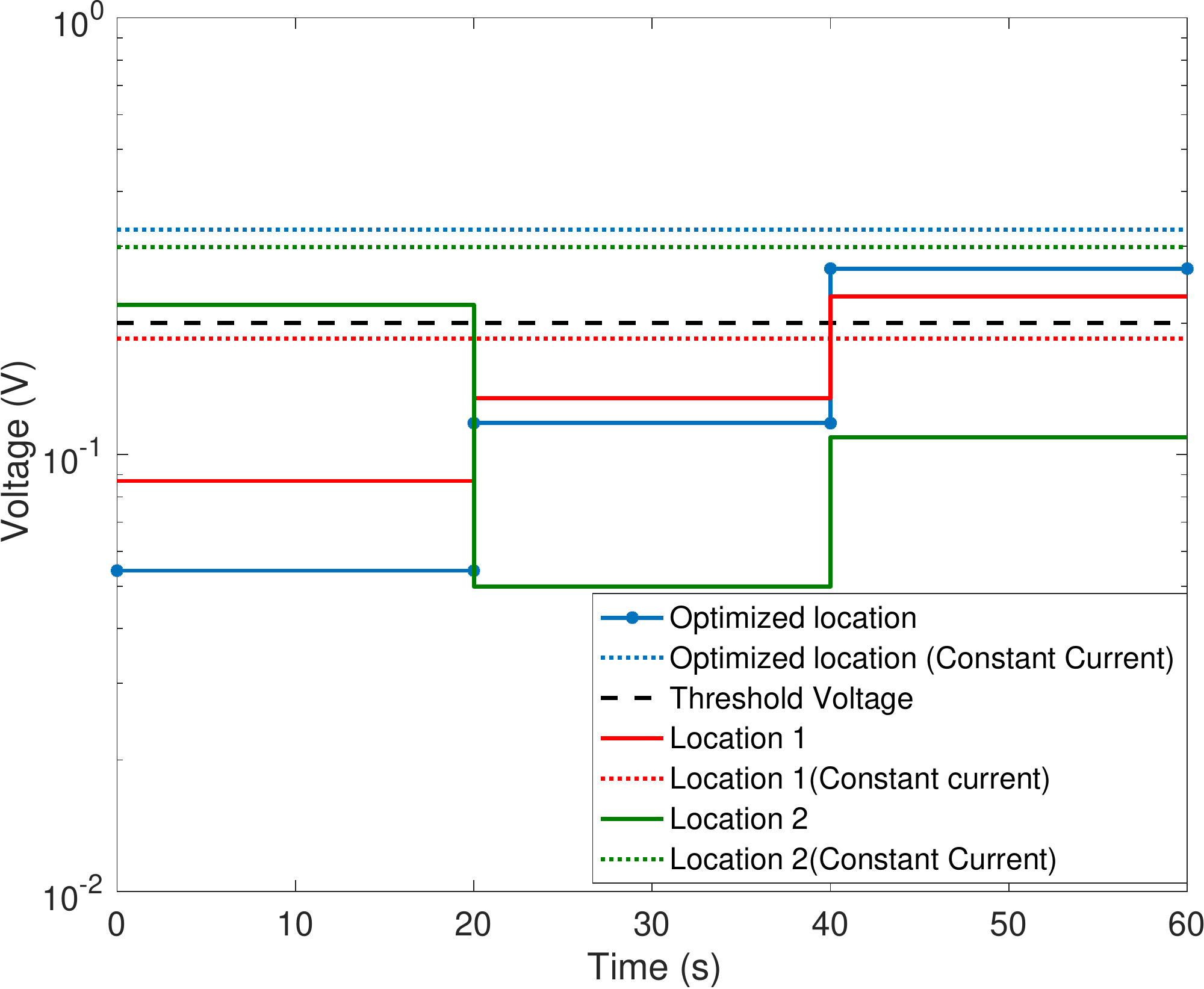}
	\caption{Illustration showing the charging performance of a current set optimized for 3 mutually orthogonal receiver orientations at the various simulation locations.}
	\label{fig:6}
\end{figure}

\begin{figure}
	\centering
	\includegraphics[width=8cm, height=6cm]{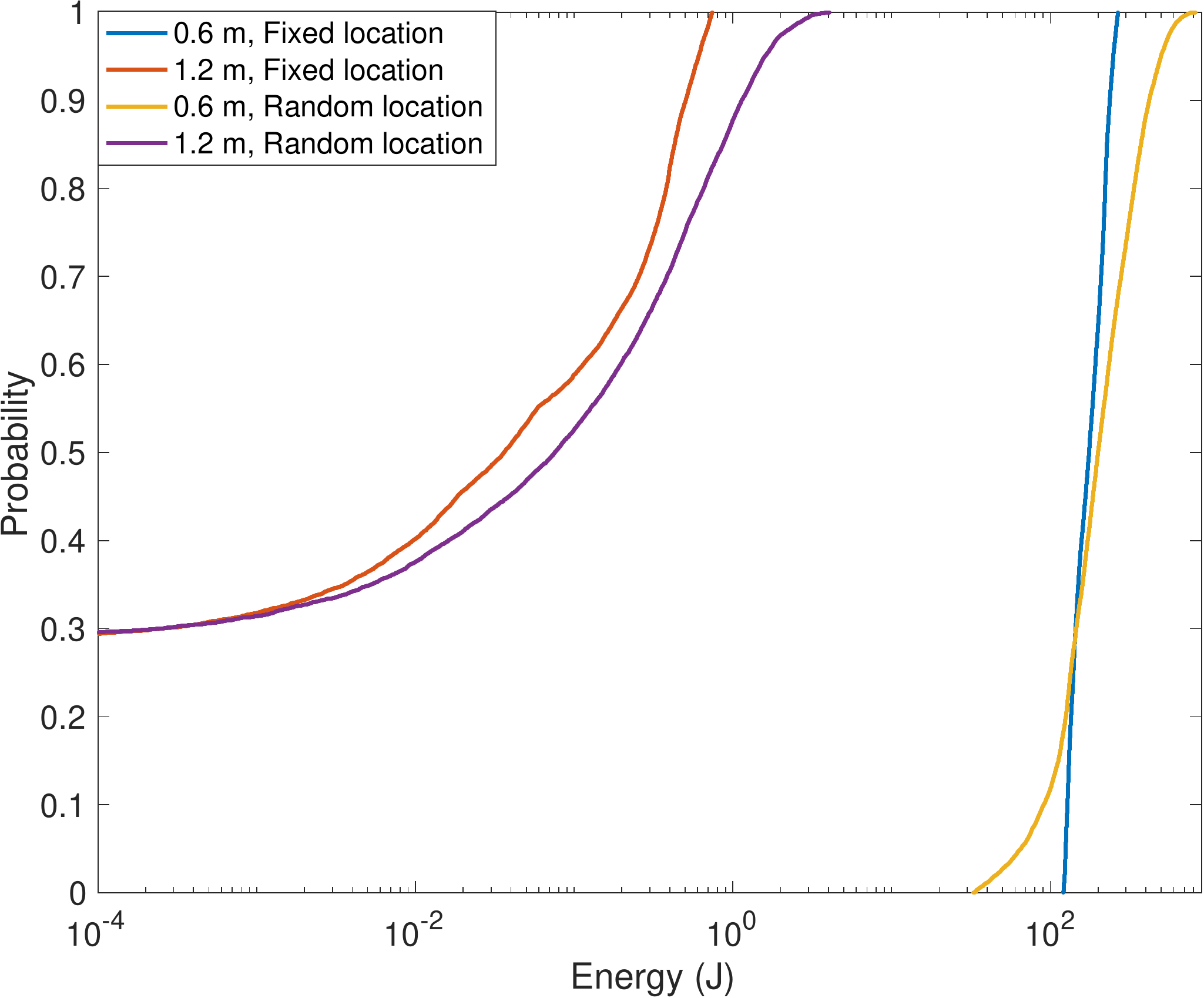}
	\caption{ CDF showing the performance of current set optimized for 3 mutually orthogonal receiver orientations at 0.6~m and 1.2~m radial distance}
	\label{fig:9}
\end{figure}

First, in Fig.~\ref{fig:8}, we fix the receiver location and randomly generate its orientation. We use 4, 8, 36, and 100 current vectors. The results show that when the distance is 1.2~m using a constant current vector, around 40\% of the receivers cannot be charged at all. If we use MagBB, the percentage decreases to nearly zero as the number of current vectors increases. For a shorter distance of 0.6~m, we have the same observation but the percentage is much lower. Also, we notice that the received energy decreases dramatically as the distance increases. This is due to the small size of the receiving coil. Since the batteryless sensors have low power consumption, the harvested energy at the level of 1 mJ can be sufficient. Also, the charging time can be increased to meet the requirement of high-power-consumption sensors. 

In Fig.~\ref{fig:7}, we vary both the receiver location and coil orientation. Our comparison suggests that there is a 60\% probability that a randomly oriented coil is not charged at all by a constant current compared to about 29\% and 23\% for a 4-current and 8-current set respectively at 1.2~m. The performance gains achieved by using the optimized current sets are more pronounced at 1.2~m than at 0.6~m. A 4-current set is therefore more computationally light and produces similar performance at 0.6~m, however, its performance deteriorates rapidly as the distance increases. At 0.6~m, there is about a 10\% probability that a randomly oriented coil will not be charged at all. Comparing Fig.~\ref{fig:8} and Fig.~\ref{fig:7}, variations in orientation at the same location have more of an effect on performance between different current sets compared to the variations in location. 

To evaluate the performance of the minimum number of current vectors, i.e., 3 current vectors, in Fig.~\ref{fig:6} we plot the induced voltages in three coils at the optimized location, location 1 and location 2. Similarly, using constant currents cannot guarantee a reliable performance; the induced voltages can be higher or lower than the threshold voltage. In Fig.~\ref{fig:9} we show the CDF of the received energy by using 3 orthogonal current vectors. Comparing with the results in Fig.~\ref{fig:8} and Fig.~\ref{fig:7}, the probability of receiving zero energy at 1.2 m is lower than using constant current but higher than using 4 or more current vectors. At 0.6 m, the performance is similar to using more current vectors. Thus, if the wireless energy transfer range is long, i.e., around 1~m, it is more reliable to use more than 8 current vectors. On the contrary, if the range is short, i.e., around 0.5~m, it is sufficient to use 3 current vectors.

\section{CONCLUSION}
A magnetic blind beamforming (MagBB) algorithm is proposed in this paper to wirelessly charge batteryless sensors with unknown coil orientations. The efficiency of magnetic induction-based wireless energy transfer strongly depends on the coil alignment. The misalignment causes significant losses. Since batteryless sensors use diodes in the energy harvesting circuit, if the induced voltage is lower than the diode threshold voltage, no energy can be received. MagBB can rotate the magnetic fields at a point which can ensure that part of the magnetic fields can be leveraged to charge a sensor. This paper provides a detailed design procedure for MagBB and discusses its reliability, location dependency, and efficiency. Extensive simulations are performed to evaluate the performance of MagBB. The results suggest that MagBB can effectively charge sensors without knowing their coil orientation. This can enable the use of large numbers of tiny batteryless sensors for applications such as precision agriculture and structural health monitoring.  

\section*{Acknowledgment}
This material is based upon work supported by the National Science Foundation under Grant No. CNS1947748. Any opinions, findings, and conclusions or recommendations expressed in this material are those of the authors and do not necessarily reflect the views of the National Science Foundation.

\bibliographystyle{IEEEtran}
\bibliography{main}

\end{document}